# Brittle to Quasi-Brittle Transition and Crack Initiation Precursors in Disordered Crystals

S. Papanikolaou[a], J. Thibault[a], C. Woodward[b], P. Shanthraj[c], F. Roters[c]

[a] Benjamin M. Statler College of Engineering, West Virginia University, Morgantown, WV 26505, USA

[b] Materials and Manufacturing Directorate, Air Force Research Laboratory, Wright Patterson Air Force Base, Dayton, OH 45433–7817, USA

[c] Max-Planck-Institut für Eisenforschung, Max-Planck-Straße 1, 40237 Düsseldorf, Germany

## Abstract

Crack initiation emerges due to a combination of elasticity, plasticity, and disorder, and it is heavily dependent on the material's microstructural details. In this paper, we investigate brittle metals with coarse-grained, microstructural disorder that could originate in a material's manufacturing process, such as alloying. As an investigational tool, we consider crack initiation from a surface, ellipsoidal notch: As the radius of curvature at the notch increases, there is a dynamic transition from notch-induced crack initiation to bulk-disorder crack nucleation. We perform extensive and realistic simulations using a phase-field approach coupled to crystal plasticity. Furthermore, the microstructural disorder and notch width are varied in order to study the transition. We identify this transition for various disorder strengths in terms of the damage evolution. Above the transition, we identify detectable precursors to crack initiation that we quantify in terms of the expected stress drops during mode I fracture loading. We discuss ways to observe and analyze this brittle to quasi-brittle transition in experiments.

## 1   Introduction

Brittle, heterogeneous materials such as rock, concrete, ice, ceramics, and various composites demonstrate complex fracture mechanics due to several characteristics such as size effects and stochasticity, leading to fractal characteristics of the fracture surface topography (Bouchaud et al. 1990, Bouchaud 1997, Kumar et al. 2011). Size effects have been well recognized as a major component of fracture: Because of the large scale use in the construction industry for dams and buildings, materials like concrete or composites are difficult to study in terms of fracture due to the size effect limitations in a laboratory (Mier et al. 1997). Size effects can be related to a length-scale that is intrinsic for the material's disorder (Bazant and Planas 1997, Alava et al. 2008): If the length-scale of the disorder is significantly greater than the nominal crack length scale, then the stress intensity at the crack tip is saturated with disorder effects. Moreover, size effects are connected to stochasticity and rare events that are otherwise statistically improbable(Alava et al., 2006). While concrete has evidently large structural disorder, generic material heterogeneity may be more subtle but analogously complex. For example, in alloys and superalloys, structural disorder is prevalent, influencing crack initiation features. Interesting examples are discussed in Ritchie and Peters (2001), where alloying disorder present in the microstructure yields crack growth rates that span several orders of magnitude for the average length of small cracks. More specifically, in the case of titanium aluminide alloy, TiAl, below the brittle-ductile transition temperature, there are crack growth rates that can vary by approximately 4 orders of magnitude. For this regime, Paris law is invalid and the stochastic nature of disorder serves as the defining rule of short crack growth (Donald and Paris 1999, Jones 2014). While nontrivially complex and stochastic, it is the short-crack limit that controls crack initiation in intermetallic alloys. Therefore, any sincere effort towards predicting fracture should be able to consistently capture this regime. Furthermore, it is a target to fundamentally understand whether short-crack growth events in brittle materials can serve as statistical precursors of fracture; the development of a set of conditions and protocols for the emergence of such precursors remains to be seen. Obviously, the observation of such precursors can permit the prognosis of crack propagation. Given the limited statistical sampling of short-crack growth, it is, also, imperative to clarify ways and protocols that can capture, controllably, this behavior. In this paper, we envision an experimental protocol for identifying and controlling the onset of such precursor events. We achieve this by utilizing the design of the curvature of a notch (Gdoutos 2005) in a pre-assumed disordered microstructure.

Crack initiation from a sharp notch is commonly believed to be controlled by pure elasticity arguments. Zehnder (2012) explores the concept of linear elastic fracture mechanics (LEFM) in great detail. LEFM focuses mainly on the macrostructural geometric characteristics at the crack tip. It allows for the simplification of a complex problem through an analytical solution and it can use parameters like stress intensity factor and displacement fields to determine if failure has occurred. However, as the notch curvature decreases, crack initiation should naturally be controlled by features of bulk disorder; the length-scale at which this happens is connected to disorder. Microstructural disorder characteristics can physically manifest themselves through heterogeneity in the material, such as the presence of other phase particles, plasticity, and inhomogeneous/anisotropic elastic properties. Microstructural disorder can be tracked through various microscopy methods such as SEM, TEM, AFM, etc. However, these observations offer no significant advancement in determining the effect on fracture behavior. In contrast, the existence of this natural transition from notch-driven crack initiation to bulk-disorder crack nucleation, as notch curvature decreases, allows for the implicit quantification of disorder. In this paper, we present a computational study of this behavior and transition.

We employ a phase field model to examine crack initiation in disordered crystals. Disorder in the model is captured through using the Weierstrass-Mandelbrot function: It produces a surface topography in D dimensions that can be used to display random patterns with controlled fractality (Aviles et al. 1987), (Power and Tullis 1991), (Borodich 1997), and (Shanthraj et al. 2011). With the introduction of the fractal dimension, a single parameter may be used to control the landscape of the parameter's fluctuations in a material where the parameter can correspond to any material property. In this model, we focus on the random, microstructurally-induced fluctuations in the energy needed to incrementally damage a representative volume element of the material. Such fluctuations may correspond to the coarse-grained effect of various flaws and imperfections on crack initiation.  Even though we use a phase field model, our approach is relatively novel; others have attempted to model the fracture behavior of small cracks, accounting for some approximation of disorder.

Computational modeling of disordered microstructures for the study of fracture has been attempted in the past using various approaches (Needleman and Tvergaard 1987, Mathur et al. 1996, Tvergaard and Needleman 2006, Needleman et al. 2012, Ponson et al. 2013, Tang et al. 2013, Srivastava et al. 2014, and Osovski et al. 2015). Two types of mechanism-based techniques for predicting fracture toughness have been debated (Gao et al. 2005). The first is to assume that the microstructural imperfections are implied using a continuum model. The Gurson-Tvergaard model follows this approach (Gurson 1977, Tvergaard 1982), allowing for the statistical modeling of the materials microstructure to be approximated. The second approach is to model the individual phase particles and voids in the microstructure of the material and implement a finite element analysis (Brocks et al. 1995, Xia et al. 1995, Gao at al. 1998a and 1998b, and Srivastava et al. 2017). However, there is a limitation on the number of voids that can be used within the model. Moreover, porosity in heterogeneous materials implies that disorder is present throughout, and stochastic behavior of fracture extends beyond individual voids and other microstructural defects. The phase field model used in this paper fundamentally assumes that crack propagation occurs in a continuum manner, modeling the statistical distribution of the material toughness on the macrostructural scale. This approach gives a simplified, but realistic fracture behavior that can track fracture precursors and consequently allow for the investigation of transitional behavior as disorder and other parameters are tuned.

Heterogeneous materials like concrete have been known to exhibit such stochastic characteristics during fracture. So much so in fact that a new classification of fracture was developed to encompass this behavior known as quasi-brittle fracture (Bazant and Planas 1998). This class of fracture behavior implies that the disorder of such materials, typically observed in the brittle fracture regime, can show significant damage accumulation before sample failure. This behavior places severe limitations on the ability to observe such events as crack nucleation. When crack nucleation occurs on the sample, the area around the crack tip is considered the fracture process zone (FPZ) where it is assumed that damage is localized to this area (Alava et al. 2008). Though disorder greatly effects the location where a crack will nucleate, the FPZ is still valid. Bazant (2004) considered that the original parameters in the FPZ such as the stress at the crack tip, $\sigma_c$, and stress intensity factor, $K_c = \sigma_c\sqrt{\pi a}$, where $a$ is the crack length, are supplemented by an additional component of disorder, altering the general equation for stress from $\sigma_c \sim K_c/\sqrt{a_0}$ to $\sigma_c \sim K_c/\sqrt{\xi + a_0}$, where $\xi$ is length-scale of disorder, $a_0$ is the initial crack length, and $K_c \sim \sqrt{EG_c}$ is a function of the material fracture toughness, $G_c$, and the modulus of elasticity, $E$. Through the use of this disorder length-scale, several important effects are noted: First, if the notch is sufficiently large, $\xi/a_0 << 0$, then the fracture behavior lies within linear-elastic fracture regime. Second, if the crack length vanishes, $a_0 \rightarrow 0$, then the average strength remains finite. Therefore, in order to observe the transition where disordered effects are significant enough to influence crack nucleation, a notch is used to drive crack initiation at the tip, given well-observed, brittle fracture behavior in the elastic regime. As the curvature of the notch decreases, the length-scale of disorder in the FPZ influences crack nucleation to occur in the bulk sample. Through the use of manipulating the curvature of the notch and the length-scale of the disorder, we develop a protocol that dictates the transition of the fracture behavior at which the material's disorder can no longer be neglected in heterogeneous materials.

Toward a simulation of this transition notch-driven crack initiation to bulk crack nucleation, we use an integrated spectral phase field approach coupled constitutively to elastoplasticity through the Düsseldorf Advanced MAterial Simulation Kit, DAMASK, software (Roters et al. 2012). This is an open-source software which we modify to introduce a microscopic roughness for the energy required for local damage. Elasticity is solved through a spectral solver which is fully parallelizable. Crystal plasticity is modeled through typical constitutive laws that take into account grain orientation, crystalline structure, and possible slip systems (Asaro 1983). Through the application of this multiscale, micromechanical analysis, the microstructure can be linked to the macroscopic material toughness energy. In this paper, we perform extensive simulations of a notched specimen in a system of dimensions: $L_x$ x $L_y$ x $L_z$. The test samples have an assumed length scale of 1 mm per 256 units (essentially, each unit corresponds to ~ 4 µm). $L_x$ and $L_z$ are held constant at 8 µm and 1.0 mm, respectively. $L_y$ is varied from 0.25 mm to 1.0 mm. The ellipsoidal notch is placed along the y-axis at the edge of the specimen where the major axis, a=$L_y$/8, is held constant. However, in order to observe the transition from notch-driven crack initiation to bulk-induced crack nucleation, the notch width, W, is varied at values from 16 µm to 0.5 mm, increasing the radius of curvature. In order to modify disorder, we consider quenched stochastic contribution in the phase field energy that controls the rate of local damage. Defining the ratio of the variance of this disorder with respect to the average phase field energy as $R_G$, we consider various cases where $R_G$ varies in the interval of [0.0, 0.8]. The stochastic variable is picked through the Weierstrass-Mandelbrot function (Shanthraj et al. 2011) in which the fractal dimension, D, is either 2.85 or 2.995. As shown in Figure 1, the sample has an induced disorder distribution that varies the phase field energy, according to the parameters and dimensions aforementioned. We are using elasticity and plasticity material properties for aluminum. In section 2, we explain how the fractal surface topography is generated and the algorithmic structure of DAMASK. Also, we discuss the phase field approach which solves for damage evolution. In section 3, we discuss the simulation results and how: a) D influences the fracture behavior, b) induced-disorder affects the location of crack nucleation, and c) the introduction of a notch affects crack initiation. Furthermore, we discuss how the transition from notch-driven crack initiation to bulk-disorder crack nucleation indicates a transition from brittle to quasi-brittle fracture regimes. Finally, in section 4, we make several important conclusions about this transition and how it is causally related to disorder towards identifying the quasi-brittle fracture regime.

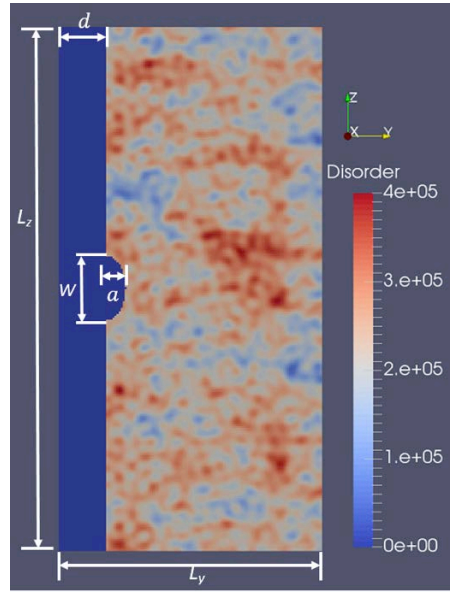

Figure 1: **Basic dimensions with an induced disorder distribution:** A disorder strength of $R_G$=0.2 was chosen where the thickness of the sample, $L_x$, is 8 µm, the width of the sample, $L_z$, is 1.0 mm, the length of the sample, $L_y$, is 0.5 mm, W is the initial crack width is 128 µm, d is 96 µm, and a is the constant major axis of 64 µm.

## 2  Fractal Surface Generation and Free Energy Behavior

DAMASK utilizes a phase field model in the continuum to develop constitutive laws towards solving for material deformation due to the damage evolution within the sample. It uses a spectral method to solve for the elastic and plastic deformation with fast Fourier transforms (FFT). Fourier methods require a rectangular-gridded mesh and periodic boundary conditions. Therefore, we apply a layer of air in contact with the notch surface, satisfying the periodic boundary conditions, in order to resolve equations for the damage evolution. DAMASK solves a material point model that converges to a solution at each time-step for various multiscale characteristics of the boundary value problem (Roters et al. 2012). In this section, we discuss: how disorder of the material is generated, the structure of the material point model, and the phase field approach.

### 2.1  Fractal Surface Topography Generation

The fractal surface topography in a heterogeneous sample is considered as the induced disorder of material bulk. The Weierstrass-Mandelbrot (W-M) function allows for modeling of such macrostructural disorder using microscopic roughness in the energy required for local damage to statistically quantify microstructural imperfections. More specifically, it creates a Gaussian distribution of this fluctuating energy (Weierstrass 1895 and Mandelbrot 1977).

$$Z(x,y) = L\left(\frac{G}{L}\right)^{D-2}\left(\frac{ln\gamma}{M}\right)^{1/2}\sum_{m=1}^{M}\sum_{n=0}^{n_{max}}\gamma^{(D-3)n}\times\left\{\cos(\phi_{m,n}) - \cos\left[\frac{2\pi\gamma^{n}(x^{2}+y^{2})^{1/2}}{L}\times cos\left(\tan^{-1}\left(\frac{y}{x}\right)-\frac{\pi m}{M}\right)+\phi_{m,n}\right]\right\}$$

(1)

Within the W-M function, the parameter, G [m], is defined as the variance of the spatial fluctuations, and the fractal dimension, D, exhibits a strong influence over the surface roughness of the function for fracture surface topography and stochasticity. Therefore, the critical parameters that cause the most evident changes in the stochasticity and effect the crack growth and damage evolution are G and D (Mandelbrot et al. 1984, Shanthraj et al. 2011, and Carney and Mecholsky 2013). However, there are other parameters that help manipulate the fracture behavior: L [m] is the sample length, $\gamma$ is a scaling parameter, M is the superposed ridges used to construct the surface for its roughness, $\phi_{m,n}$ is a random phase with m and n defining the ridge and frequency index, n is a frequency index that must control the max index considered, Z(x, y) [m] is the surface function, and x [m] and y [m] correspond to locations in the Fourier grid. The W-M function is used to manipulate the surface roughness distribution of the sample and provides a statistical disorder distribution in order to simulate fracture behavior.

## 2.2 Phase Field Approach

In this work, we use DAMASK (Roters et al. 2012) to investigate crack initiation and growth in brittle disordered alloys from an engineered notch. DAMASK uses typical constitutive elasticity modeling and we also model plasticity through the popular constitutive rate dependent, but also slip dependent, crystal plasticity formulation (Asaro and Lumbarda, 2006). The structure of the algorithm is using typical material point modeling. The structure of the material point model begins by examining the periodic boundary conditions of the sample. Using the spectral method, an FE analysis is used and the FFT determines the average deformation gradient, $\bar{F}$. Solving for the average deformation gradient requires an iterative analysis to converge on a solution for each incremental time step. However, it requires the corresponding average first Piola-Kirchhoff stress, $\bar{P}$, at each mesh point. Through partitioning the individual deformation gradients, the first Piola-Kirchhoff stress information is used to solve for the stress within the crystalline structure. Then, the second Piola-Kirchhoff stresses, S, are used to find the plastic velocity gradient, $L_p$, in order to solve the elasto-plasticity problem in the crystal..(Roters et al. 2012). This approach to study quasibrittle behavior in disordered alloys is a true leap compared to recent efforts on identifying stochastic effects in fracture (Shekhawat et al. 2013). While this work focuses on the effects of curvature on quasi-brittle behavior, it is a concrete aim to perform scaling analyses in this highly realistic and experimentally comparable setting.

For the investigation of cracks, DAMASK utilizes a clear phase field approach (Aranson et al. 2000) where cracks are modeled by a phase with order parameter $\phi=0$, while the material has $\phi=1$. $\phi$ couples proportionally to the elastic coefficients of the material, softening as it is damaged. DAMASK utilizes a spectral method to resolve the elastic, plastic and damage equations during crack growth. These equations originate in Griffith's Energy criterion as there are three types of free energy during fracture of the sample. Elastic free energy is quite prevalent within the sample as it is loaded. When the strain energy release rate reaches a critical point the elastic free energy in the disordered distribution, locally, is converted to plastic free energy. The critical strain energy release rate is how the phase field models the disorder and determines how fracture behavior occurs in the model. As loading continues, the plastic deformation occurs and micro-cracks begin to present as damage in the sample. Due to the damage evolution in the disordered specimens that are the focus of this work, a crack will either initiate at the notch due to LEFM or within the bulk-disorder of the sample due to the compromised macrostructural setting.

The damage accumulation is determined by the Fourier grid variables as the material toughness energy values are assigned. As the strain is increased, the mesh is used to track all of the respective energy contributions at each incrementally-increasing strain-steps. With this progression, there is deformation in the sample that can be specified. However, due to the complexity of modeling heterogeneity and anomalous fracture behavior, more general relationships for Griffith's energy criterion are used. The thermodynamically consistent free energy form can be used to determine the deformation.

$$\psi = \Psi(\nabla X, F_p, \xi, \phi, \nabla\phi)$$

(2)

Where $\psi$ is the collective free energy of a volume element, $\nabla X$ is the total deformation gradient, $F_p$ is the plastic deformation gradient, $\phi$ is the phase field, and the $\nabla\phi$ is the phase field gradient. Equation 2 expresses the most general form of the relationship for the free energy equation. We further make the basic hypothesis that the elastic energy is decomposed in a direct multiplicative manner to the simplest damage contribution ($\phi^2$) and a pure distortion-dependent contribution:

$$\psi_E = \phi^2\widetilde{\psi_E}(\nabla X, F_p)$$

(3)

At the undamaged state where crack initiation hasn't occurred, the energy contribution to the sample is elastic free energy.

In equation 3, it is a function of the stored elastic energy density, $\widetilde{\psi_E}$, and the phase field. The stored elastic energy density is dependent on the product of the total deformation gradient and plastic deformations. As the loading progresses, damage accumulation occurs at a location in the bulk sample that exhibit the lowest material toughness energy. Therefore, at crack initiation, the elastic free energy is zero, locally, at the crack tip (Shanthraj et al. 2016) and the only energy present are plastic and damage energy.

Analysis of the plastic free energy is much more complex. It requires at set of equations that represent several different contributions that affect plastic energy and physically it only occurs at the crack initiation and propagation areas. The plastic deformation gradient must be considered first.

$$\partial_{F_p}\psi = -\phi^2 S F^{-T}$$

(4)

Where S is the second Piola-Kirchhoff stress in terms of,

$$S = C(F_e^T F_e - I)/2$$

(5)

Also, modeling the plastic velocity gradient $L_p$ by means of the plastic slip rate,$\dot{\gamma}$ systems.

$$L_p = \dot{F}_p F_p^{-1} = \sum_\alpha \dot{\gamma}^\alpha s^\alpha \otimes n^\alpha$$

(6)

where s and n are characteristic of the dislocation slip system. However, the slip rate is guided by the basic phenomenological crystal plasticity constitutive equation.

$$\dot{\gamma}^\alpha = \dot{\gamma}_o \left|\frac{\tau^\alpha}{g^\alpha}\right|^n sgn(\tau^\alpha)$$

(7)

In our simulations, we consider single crystals at fixed crystalline orientation, and all 12 FCC slip systems may in principle become active. When the plastic energy reaches its critical value, the energy is converted to damage free energy. The damage free energy is nonzero only after the crack initiation event,

$$\psi_D = \frac{1}{2} G_c l |\nabla\phi|^2 + \frac{G_c}{l}(1-\phi)^m + I_{[0,1]}(\phi)$$

(8)

Equation 8 represents how the damage free energy is determined. There are two critical parameters: $G_c$ and $l$. $G_c$ is typically labeled as the critical strain energy release rate, analogous to the work of fracture. In our field model, this amounts to the free energy contribution necessary for crack nucleation, and $l$ is the resolution length scale which should correspond to a representative volume element (RVE) of the material, that satisfies mechanical equilibrium (Shanthraj et al. 2016). In our simulations, we consider a fixed resolution length equal to 4 mesh units, which corresponds to ~16 μm, given our considered sample dimensions.

In this work, we identify disorder in terms of the variability for the critical strain energy release rate $G_c$. We assume throughout that $G_c$ displays local, quenched and continuous fluctuations, as it would naturally happen in a multi-phased disordered metal alloy or a fatigued specimen. Namely, we assume that $G_c \rightarrow G_c + \Delta G_c(r)$ and the local fluctuations $\Delta G_c(r)$ are given in space through the aforementioned W-M function at a given fractal dimension D. The critical experimentally relevant parameter in these simulations is the degree of relative disorder fluctuations, or in other words the width of the distribution of $G_c$ with respect to the average critical strain energy release rate. If we define the standard deviation of the distribution as $\delta G_c$, then the important quantity to investigate would be $R_G = \delta G_c / G_c$. In the next section, we discuss how this phase field model in a realistic elastoplastic environment can cause crack initiation and fracture at various notch widths and disorder strengths, keeping $G_c$ fixed but changing the disorder ratio $R_G$.

## 3   Simulation Results

The objective of the simulations is to identify the characteristic effect of the stochastic local damage distributions and how they contribute to the fracture mechanics of the specimen. When crack initiation occurs, the natural outcome is to occur at the notch, given that the stress intensity factor allows for far higher stress at the notch tip than any place in the material, allowing failure to occur at this point. However, the induced disorder distribution in the specimen could allow for crack nucleation away from the tip of the notch.

### 3.1   Effects of Fractal Dimension D on Fracture Behavior

In order to increase the variation or fluctuation in the local critical strain energy release rate, $\delta G_c$, that naturally competes with the discretization unit of the simulation that is used for modeling the notch curvature, the fractal dimension

D is adjusted. The fractal dimension is increased to analyze the character of crack propagation as a result of the change. The fractal dimension is defined in the W-M Function and plays an important role in the profile of the surface roughness. The fractal parameter, also known as the Hausdorff-Besicovitch dimension, adds more resolution to the surface roughness. The main goal of increasing the fractal dimension of the material's damage parameter throughout the sample is to produce much more defined crack initiation and propagation profiles. Simulations were conducted at several fractal dimension values to determine how the fluctuations in the critical strain energy release rate would affect the crack nucleation and propagation behavior. With the increase in fractal dimension, the crack nucleates away from the notch at lower disorder strengths because the critical strain energy release rate varies at a higher degree for lower disorder strengths compared to the lower fractal dimension.

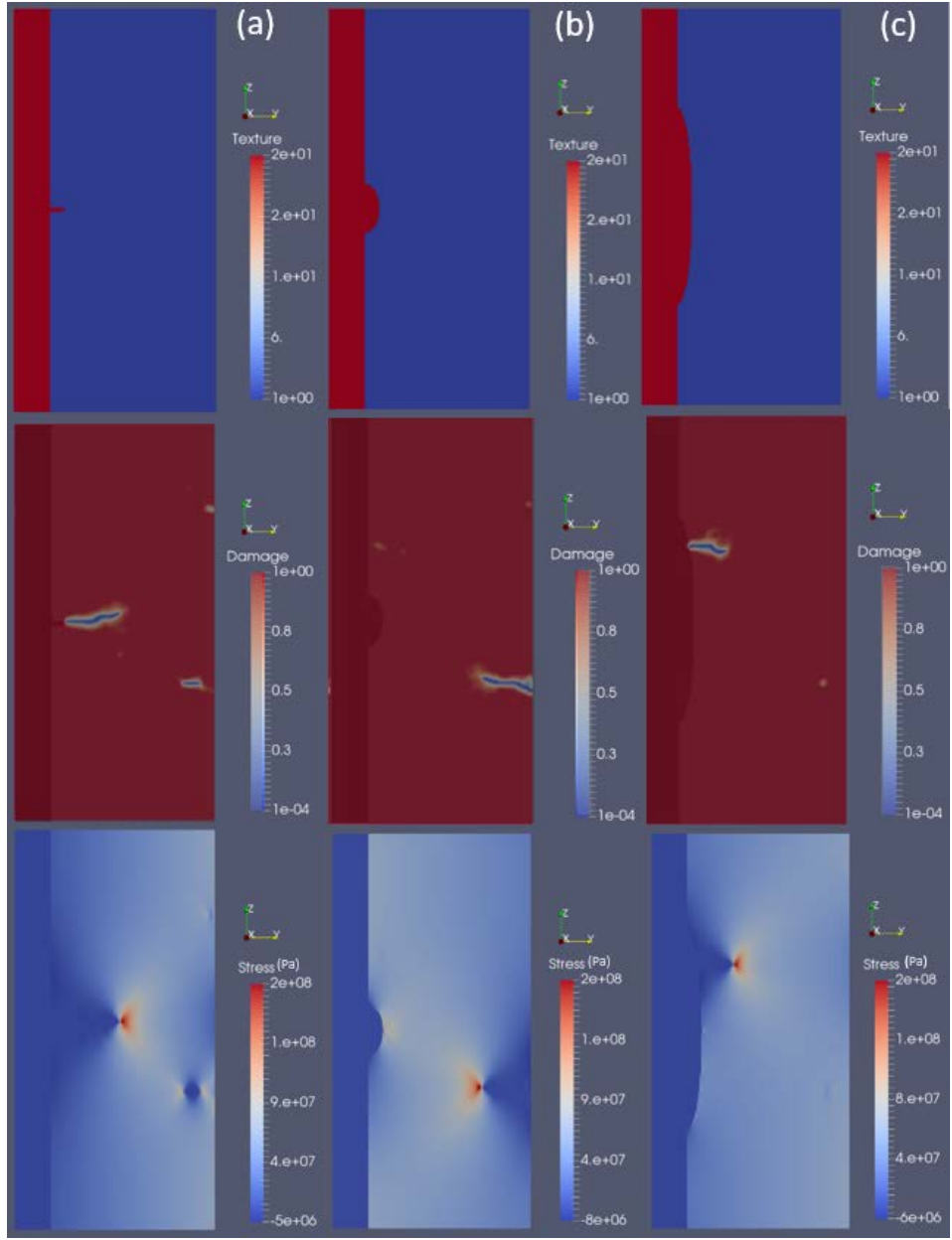

*Figure 2: **Effect of increasing notch width on crack nucleation for fractal dimension D=2.85**: Simulations of system size of $L_x$=8 μm, $L_y$= 0.5 mm, and $L_z$=1.0 mm where D=2.85, $R_G$ = 0.8, and (a) has a W of 16 μm (b) has a W of 128 μm (c) has a W of 0.5 mm where the texture, damage and stress distributions are displayed, respectively.*

Figure 2 shows the damage (where 1 is undamaged and 0 is fully damaged) and stress (Pa) distributions for initial crack widths of 16 μm, 128 μm, and 0.5 mm. Though there is a transition from notch-located crack initiation to bulk material crack nucleation, this behavior occurred at a disorder ratio $R_G$ of 0.8 where the material fluctuations are the largest on the tested interval. When the fractal surface roughness is increased, this fracture behavior occurs at disorder ratios of approximately $\sim 0.2$, much lower in the tested interval. At a disorder of 0.8 for the higher fractal dimension, the specimens produce multiple nucleation points with larger values of damage. Nevertheless, at lower fractal dimension D = 2.85, the crack propagation is quite brittle with almost no branches of secondary cracks form.

The characteristic effect that the fractal dimension has on the disorder distribution is it allows for larger quenched fluctuations in the critical strain energy release rate. In turn, secondary crack formation can be seen as fractal dimension is increased. Moreover, due to the fact that the large fractal dimension causes spatially sharper extremes in the disorder distribution, damage is much more prevalent in areas away from the notch (Fig.2). In general, it is natural to expect that the simulation results are overall independent of the damage distribution's fractal dimension, and for this reason we will be only studying two cases throughout this work (D=2.85 and 2.995). However, the fractal dimension is also crucial in bypassing the mesh discretization resolution, which naturally leads to artificial stress concentrations near the engineered ellipsoidal notch.

### 3.2 Crack initiation for fixed notch width as function of increasing damage distribution variance

Analogously to other works (Bazant et al. 1998) and (Alava et al. 2008), the increase of disorder leads to the onset of quasi-brittle behavior, namely crack nucleation initiates at bulk locations, where the weakest sites exist. Here, we investigate this phenomenon in the presence of a notch, and in the context of our simulation software DAMASK and through the use of the W-M function towards the disorder distribution generation.

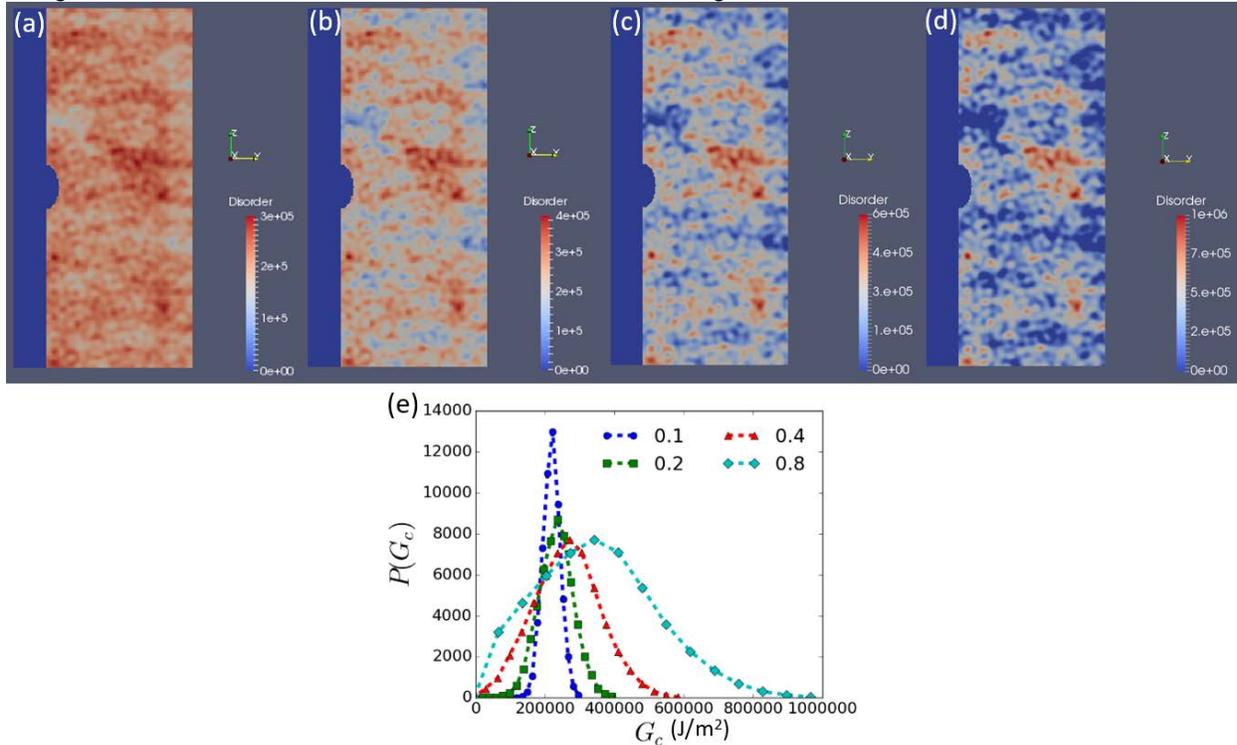

*Figure 3:* **Disorder distribution for local fracture tendency and spatial profiles::** *The progression of the disorder ratio, $R_G$, for $W = 0.125$ mm with a system size of $L_x = 8$ µm, $L_y = 0.5$ mm, and $L_z = 1.0$ mm where (a) has a disorder ratio of 0.1, (b) has a disorder ratio of 0.2, (c) has a disorder ratio of 0.4, (d) has a disorder ratio of 0.8, (e) is a histogram of the amount of fluctuating phase field energy parameter of each node vs. the phase field energy parameter.*

Figure 3 depicts a progression of the increasing $R_G$. As the disorder distributions are created at lower disorder ratios, the fluctuations in the critical J-integral value, $J_{Ic}$, will be relatively small. But at higher disorder ratios, the variation in the critical strain energy or material fracture toughness will be much higher.

Physically, the $R_G$ value allows one to control the variation in the toughness of the material. We choose the mean critical strain energy parameter, $G_c$, of the material to be a value of 200 kJ/m², rendering the sample quite brittle, more brittle than steel alloys. If the disorder distribution ratio $R_G$ is 0.1, then the variance in the surface toughness occurs in the interval of [1e5, 3e5] J/m², so the maximum and minimum toughness can only be 300 kJ/m² and 100 kJ/m², respectively. This is why, in figure 2, the distribution for $R_G = 0.1$ and W = 0.125 mm only show shades of red. However, if the disorder distribution is $R_G = 0.8$, then the variance in the surface toughness is expected to lie on the interval [0.0e0, 1.0e6] J/m²so the maximum and minimum toughness can be 1 MJ/m² and 0 J/m², respectively. This is why, in figure 2, the distribution for $R_G$ = 0.8 and W = 0.125 mm exhibits a much larger variation in the range with parts of the distribution having higher values of critical strain energy release rate, indicated by red, versus lower values of critical strain energy release rate, indicated by shades of blue. As $R_G$ increases, the legend's scale maximum is expected to increase as higher values in critical strain energy or fracture energy will be present in the distribution. Furthermore, panel (e) refers to the histogram of the critical strain energy release rate values. Higher $R_G$ curves appear to be normally distributed close to the defined mean critical strain energy release rate value aforementioned.

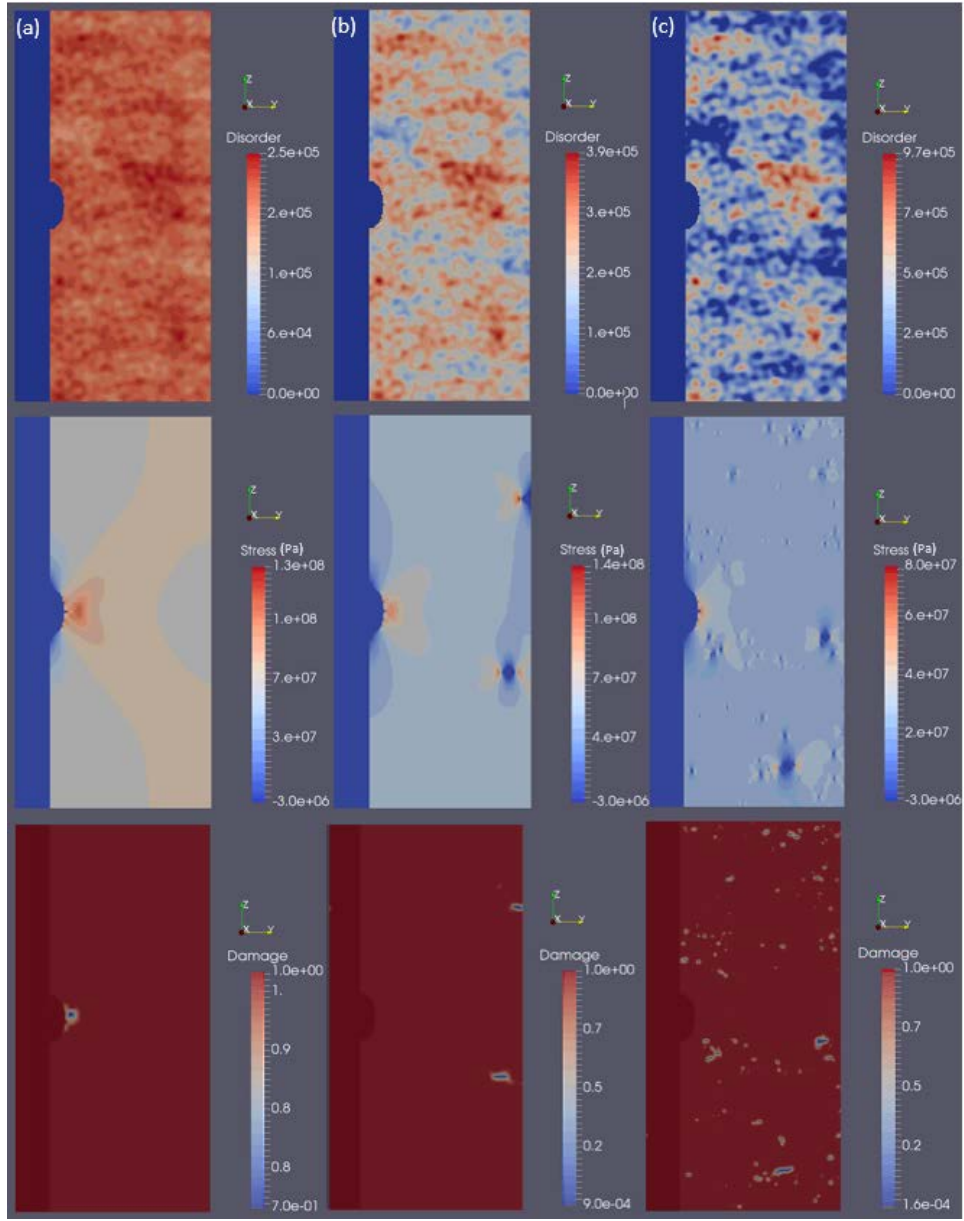

*Figure 4:* **Effect of increasing disorder strength on crack nucleation:** *Simulations of system size of $L_x$=8 µm, $L_y$=0.5 mm, and $L_z$=1.0 mm where D=2.995, W = 0.125 mm, and (a) has a $R_G$ of 0.05 (b) has a $R_G$ of 0.2 (c) has a $R_G$ of 0.8 where the disorder, stress and damage distributions are displayed (top, middle, bottom respectively), respectively.*

In Figure 4, crack initiation as function of disorder is investigated. The crack initiation occurred at the initial crack tip when the $R_G$ value is 0.05 (panel a), while the crack initiated at another location away from the notch tip for $R_G$ of 0.2 and 0.8 from (panels b and c), respectively. It can be seen that the crack propagation was affected by the increased variance in the critical strain energy release rate. At the lower disorder ratio distributions, the crack is rather straight and remains so as it propagates through the specimen. However, as the disorder ratio increases, the observed trend shows that crack nucleation occurs away from the initial notch. Then, at even higher disorder strengths, multiple cracks begin to nucleate across the entire specimen. For the cases where $R_G$ < 0.2, we observe that the crack propagates at the notch, but as it continues through the specimen, the crack's path is altered, moving through areas of lower critical strain energy release rate as a result of the introduced disorder. There is a transition in the behavior: At low disorder ratio $R_G$, the specimens' exhibit behavior that follows brittle fracture mechanics, nucleating at the notch tip, but, at higher disorder strengths, notch-driven fracture transitions from brittle to a quasi-brittle, disorder-driven fracture with nucleation behavior that is influenced by crystal plasticity.

As we focus closer on this phenomenon with respect to disorder, we examine the damage and stress distributions. However, it is pertinent to this discussion to look at the time of the damage accumulation that causes crack initiation by considering various snapshots during the loading of the specimen for the case of W = 0.125 mm and $R_G$ = 0.8.

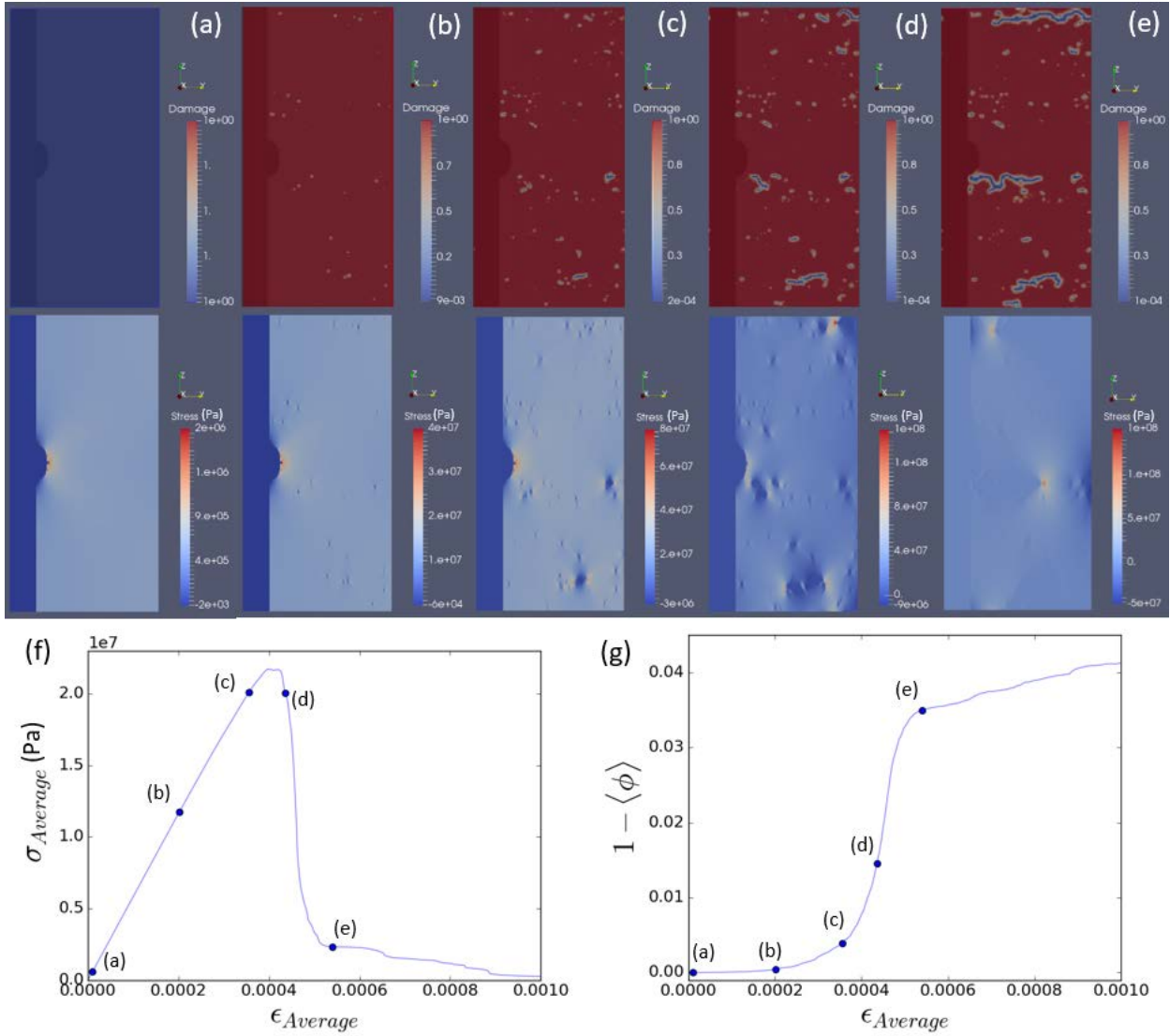

*Figure 5: **Stress and damage progression in a disordered, notched, crystalline sample:** Simulations of system size of $L_x$=8 μm, $L_y$=0.5 mm, and $L_z$=1.0 mm where D=2.995, W = 0.125 mm, and $R_G$ = 0.8 where (a) is the undamaged phase, (b) is the early damage phase, (c) is the late damage accumulation phase, (d) is the multiple crack nucleation event, and (e) is the crack propagation where the damage and stress distributions and are displayed (top, middle), respectively, and (f) and (g) are the average stress (Pa) vs. strain and average damage vs. strain, (bottom) respectively, where the five locations are identified on the curves.*

Figure 5 exhibits the fracture behavior as the load case progresses. The undamaged phase, (a), is used as a comparison with the progression of damage and stress distributions. During the undamaged phase, the specimen shows no damage.

However, there is a stress concentration at the tip of the notch, as expected, indicating that this location has the highest stress values. As the progression continues, the next phase, (b), reveals that damage has occurred in the bulk of the sample and the stress has increased at the notch tip. Stress is beginning to increase at multiple locations throughout the sample where the critical strain energy release rate is lower compared to the mean. As loading further continues, damage accumulates and crack initiation occurs, but the crack initiation occurs at multiple locations on the sample, as can be seen in panels (c), (d), and (e). There are several factors that determine the location of the inclusion that propagates through the specimen. It is determined by material properties like critical strain energy release rate, stress distribution at the current load step, statistical modelling, and proximity to other inclusions.

### 3.3 Width of Notch and its Effects on Crack Initiation and Propagation

The principle theory that guides crack initiation of a notched specimen is fracture mechanics. For our simulations, we analyze the behavior under Mode I fracture where tension or compression is used to initiate a crack. Linear Elastic Fracture Mechanics (LEFM) is used to approximate the stresses at the crack tip. However, it is well recognized that this model is inaccurate in a limited region of the Fracture Process Zone (FPZ) (Bazant 2004). The small scale yielding model for the size of this region can be approximated so that the stresses outside the process zone are accurate. For an ellipsoidal crack shape with initial crack length a, the stress intensity factor at the tip can be approximated by equation 9. But, the stress intensity factor changes everywhere in the specimen as the crack continues to grow.

$$K_I = \sigma_\infty \sqrt{\pi a}$$

(9)

Where $\sigma_\infty$ is the tensile stress that the sample is subjected to. As the length of the crack increases, the stress intensity factor at the crack tip will increase. However, crack propagation only occurs in the event that stress intensity factor, $K_I$, is greater than the critical stress intensity factor, $K_{IC}$ (Zehnder 2012). Due to the existence of spatial disorder in the elastic energy release rate, crack initiation occurs within the material bulk due to lower fracture toughness values compared to the notch tip.

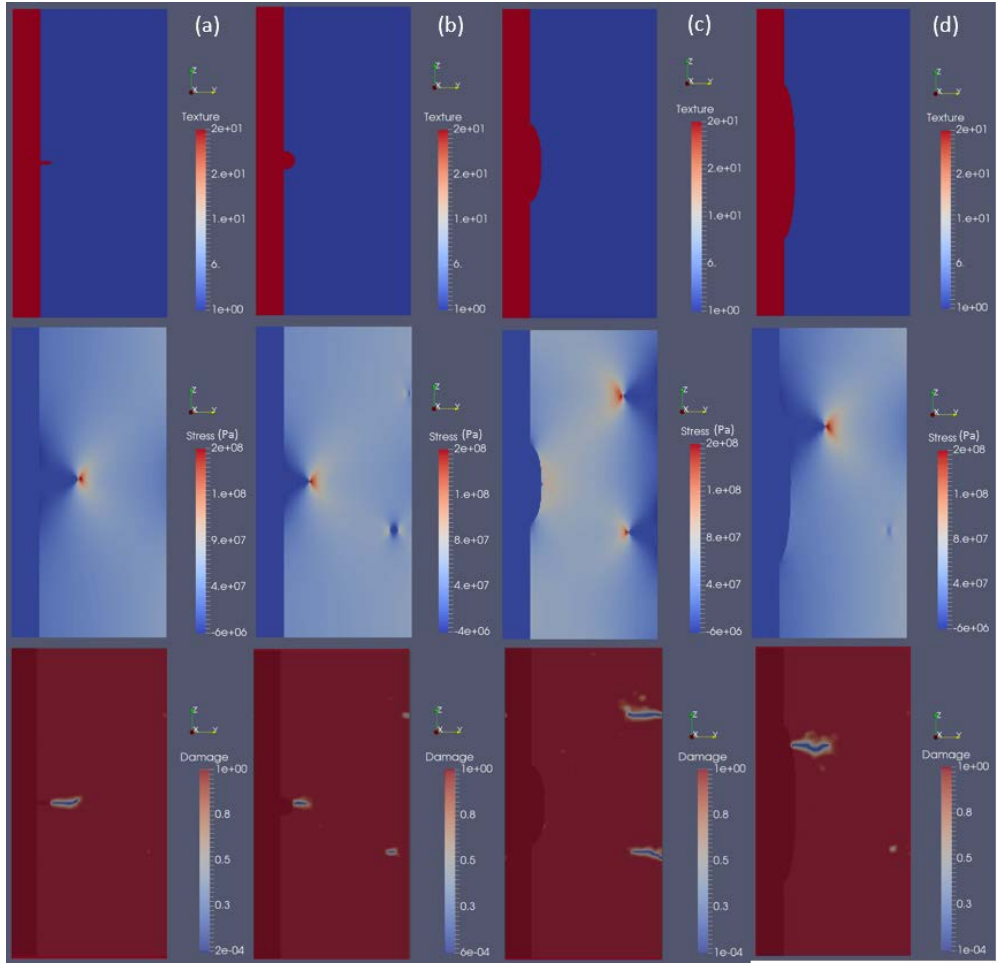

*Figure 6:* **Effect of notch width on crack initiation in a disordered crystal::** *Simulations of system size of $L_x$=8 μm,*

*$L_y$=0.5 mm, and $L_z$=1.0 mm where D=2.995, $R_G$ = 0.2, and (a) has a W of 16 µm (b) has a W of 64 µm (c) has a W of 0.25 mm (d) has a W of 0.5 mm where the texture, stress and damage distributions are displayed (top, middle, bottom), respectively.*

Figure 6 shows the crack propagation behavior of disorder strength, $R_G$ = 0.2, with respect to the various crack widths of 16 µm, 64 µm, 0.25 mm, and 0.5 mm. At relatively low disorder strengths and low notch widths of 64 µm or less, crack initiation occurs at the notch, but as W increases the location of crack nucleation will deviate from this behavior, occurring within the sample. The load case and disorder distributions for these samples were held constant across the simulations. However, there is varying behavior in the crack nucleation location and, as a result, the crack propagation behavior. As explained above, the low crack widths, like (a) and (b), exhibit large stress concentration values at the tip of the notch as the radius of curvature at the crack tip becomes smaller. Upon examination of (c) and (d), the radius of curvature is much larger, allowing for lower stresses at the notch tip compared to the rest of the stresses present in the sample. Instead of crack initiation at the notch tip, nucleation occurs at a location where the critical strain energy release rate in the phase field energy distribution is statistically lower than the critical strain energy release rate at the notch tip. This behavior is observed across the same disorder strength with increasing notch width, and shows that there is a transition from brittle, notch-driven fracture to quasi-brittle, disorder-induced fracture. As a result, the initial notch width in the specimen can be concluded to have a characteristic effect on the crack initiation and propagation behavior.

LEFM yields a predictive behavior of the crack initiation and propagation that can be solved analytically. Figure 6 shows similar characteristics because, when lower disorder strengths are considered, the critical strain energy release rate's fluctuation contribution to the propagation behavior is relatively low once crack initiation occurs at the tip of the notch. However, when considering manufacturing and processing, using alloying, of multiple components, a sample will have varying material properties which impose characteristics that influence the heterogeneity of the bulk material.

### 3.4    Average Stress and Damage Through the transition from Brittle to Quasi-Brittle Behavior

The transition in the fracture behavior is observed in terms of the average and maximum stresses, average damage, crack initiation location and crack propagation speed. It is important to note that the stress at the notch tip, in the absence of disorder, is (Inglis, 1913):

$$\sigma_{max} = \sigma_\infty \left(1 + 2\sqrt{a/\rho}\right)$$

$$(10)$$

Where the stress at the tip of the notch, $\sigma_{max}$, is a function of the applied stress, $\sigma_\infty$, the radius of curvature, $\rho$, and initial crack length, $a$. The applied stress is the tensile stress response of the sample as it is incrementally loaded at the specified strain rate. The radius of curvature, $\rho = b^2/a$, is a function of the initial crack length and the width of the notch, $W = 2b$, given that the minor axis of an ellipse is 2b. Inglis' solution expands on the previous work done by Kirsch where he analyzed the solution for a circular notch in the center of a specimen and found the stress around it to be proportionally $\sigma_{max} \sim 3\sigma_\infty$ (Kirsch 1898). Thus, the stress at the tip of the ellipse is proportional to the radius of curvature at the tip. Several observations are made from Inglis' solution: If the curvature is high, then higher stresses will be present at the tip of the ellipse. As the minor axis parameter, b, goes to zero, the stress at the tip will reach infinity. Further, the consideration of a disorder distribution would have called for a nonlinear solution, making it impossible to solve analytically. Our simulations incrementally increase the strain at a controlled rate. The stress within the material is growing until it reaches the maximum allowable stress, locally. Because of the introduction of a disorder distribution, the maximum allowable stress at any point on the sample is not uniform and the damage or crack can nucleate anywhere in the sample with a low enough critical strain energy release rate $G_c$. As the damage increases and the crack nucleates, the average stress, maximum stress, and average damage of every element in the mesh are recorded and the information is processed as the propagation occurs. The sample averages are plotted versus the average strain.

The maximum stress in the simulations is an important metric when introducing fluctuations in the critical strain energy release rate because under ideal conditions the maximum stress should be the principal driver for crack initiation.

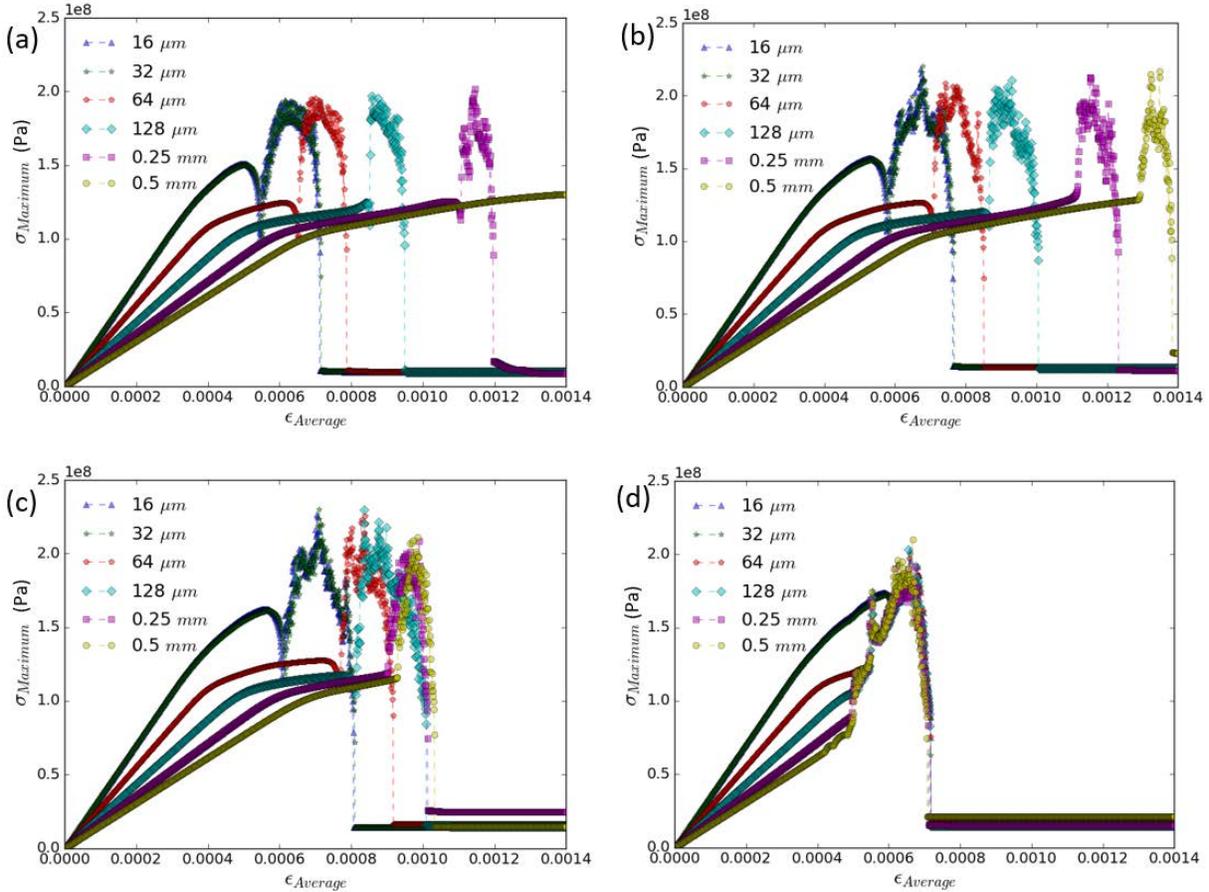

*Figure 7:* **Effect of disorder on maximum stress:** *The maximum stress of samples of $L_x$=8 µm, $L_y$=0.5 mm, and $L_z$=1.0 mm where D=2.995 as a function of the average strain with respect to $R_G$ where (a) has a $R_G$ = 0.0, (b) has a $R_G$ = 0.1, (c) has a $R_G$ = 0.2, and (d) has a $R_G$ = 0.4.*

In figure 7, at lower disorder strengths, the slope of the initial curves of maximum stress show a steady increase. At the point of crack nucleation, there is an event in which the stress decreases sharply at the time of nucleation. This can be explained as the load is increased and nucleation occurs, the stress and energy in the specimen are transferred into other forms of energy. However, as can be seen in the stress progression of figure 5, once crack initiation occurs, the stress distribution across the sample is lowered and the greatest maximum stress is at the tip of the crack because the radius of curvature is smallest at this point. As deformation continues, the maximum stress continues to obtain high values at the crack tip, but after complete failure, the maximum stress decreases sharply to a residual value. At higher disorder strengths, the maximum stress curves fail to exhibit any of the same trends that the lower disorder strength curves do. The maximum stress values across all notch widths at $R_G$ = 0.2 and $R_G$ = 0.4 show larger variation in the peak maximum stress values not consistent with the lower disorder strengths. Furthermore, the initial maximum stress decrease at crack initiation dampens out as notch width and disorder strength increase.

At low lower disorder strengths, the average stress follows the typical brittle fracture regime with no avalanches or fluctuation in the decreasing average stress. Also, the average stress required for fracture, increases as the notch width increases: Due to the decreasing radius of curvature at the notch tip, the sample requires a larger applied tensile load for damage to occur. At higher disorder strengths, the average stress enters this quasi-brittle fracture regime where there are avalanches as the sample reaches its peak average stress and crack nucleation occurs. Furthermore, this regime, also, produces fluctuations as crack propagation occurs. We focus on the behavior near the brittle-quasibrittle transition.

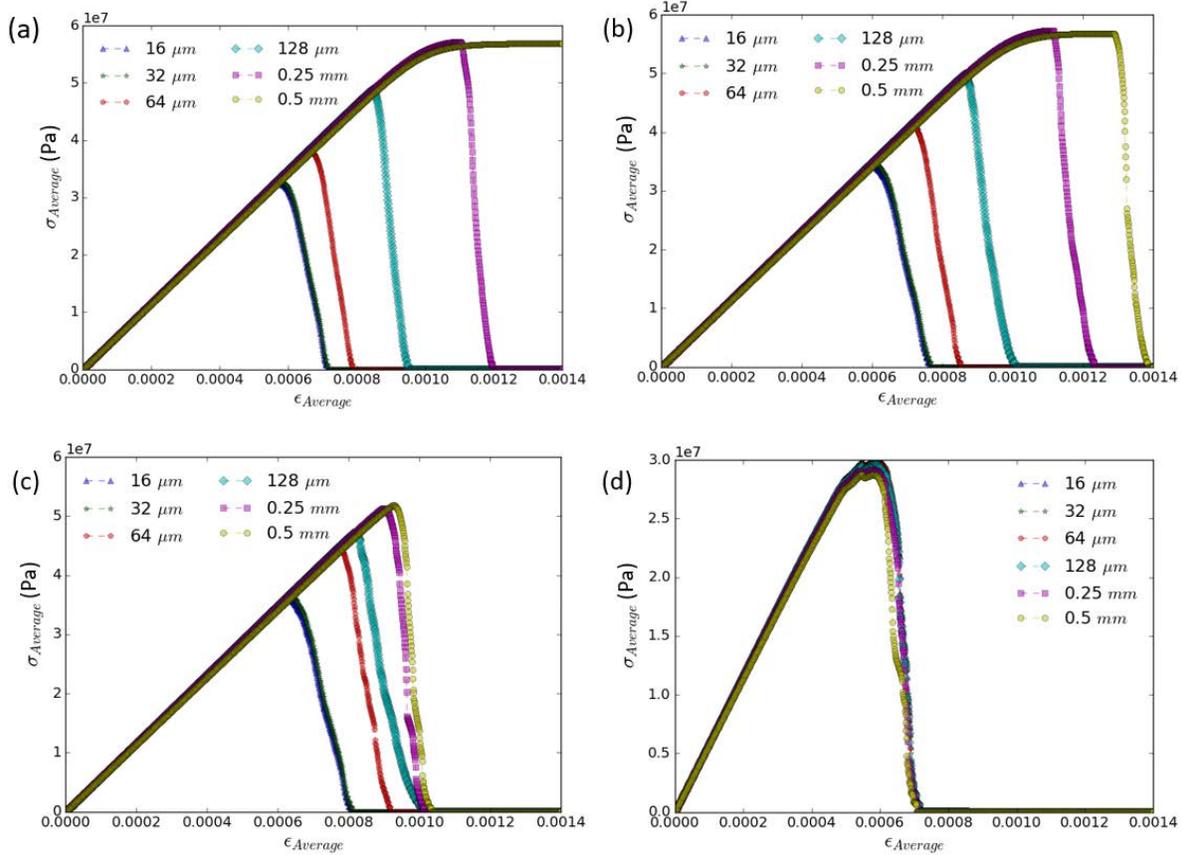

*Figure 8:* ***Effect of disorder on average stress:*** *The average stress of the specimens of $L_x$=8 µm, $L_y$=0.5 mm, and $L_z$=1.0 mm where D=2.995 as a function of the average strain with respect to $R_G$ where (a) has a $R_G$ = 0.0, (b) has a $R_G$ = 0.1, (c) has a $R_G$ = 0.2, and (d) has a $R_G$ = 0.4.*

In Figure 8, the behavior of the samples can be characterized as brittle when observing the average stress curves at the lower disorder strengths from (a) and (b). The crack initiates at the notch tip when the local critical strain energy release rate is reached and average stress begins to decrease. The average stress curve's response to this event is that the stress decreases relatively quickly, giving brittle signatures. At lower disorder strengths, the sample obtains a small, residual average stress value after complete failure and follows the expected trend: As notch width increases, the maximum average stress increases before crack initiation occurs at the notch tip. With the increasing radius of curvature, the sample exhibits a plastic characteristic: As the strain is increased, there is no increase with respect to the average stress. However, as the disorder strength increases, the maximum average stress values decrease for higher crack widths and crack initiation occurs earlier in the load case. The fluctuation at which this response occurs is of importance to classifying the fracture behavior. Furthermore, the curves appear to collapse on each other, displaying similar quasi-brittle fracture behavior.

We identify the brittle to quasi-brittle transition as the disorder strength at which there is a clearly observed insensitivity of the maximum average stress before fracture at the notch width. With the quasi-brittle fracture behavior seen at higher disorder, the samples appear to transition from notch-driven to disorder-driven crack initiation around a disorder strength of 0.2. Therefore, we introduce a phase diagram that explicitly shows where the transition from notch-driven crack initiation to bulk-disorder crack nucleation occurs. Higher disorder strengths cause crack nucleation to occur at areas in the sample with lower critical strain energy release rate relative to the overall distribution, and lower ones allow for crack initiation to occur at the notch tip. The next consideration is to identify the "critical" disorder strength where both events approximately occur simultaneously. The critical disorder strength is defined as the disorder strength ratio at which damage occurs at both the notch tip and within the bulk of the sample, and there is an observed collapse of the damage-strain curves. Through the competition of damage in the bulk material and notch tip, a curve can be approximated where, above the curve, damage occurs in the bulk first and, below the curve, damage occurs at the notch tip.

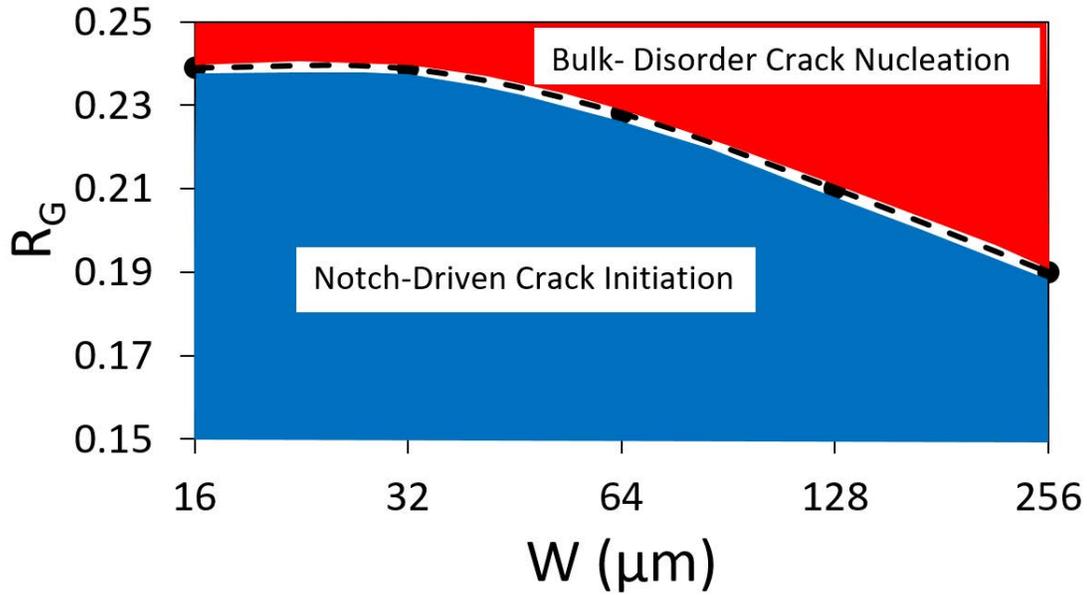

*Figure 9:* **Phase diagram of disordered brittle alloys:** *Critical disorder ratio curve for predicting the occurrence of damage at the notch width and a location of lower critical strain energy release rate in the disordered sample.*

Figure 9 shows the relationship through a phase diagram where the critical disorder strength curve is plotted with respect to crack width. The overall trend for the critical disorder curve: As crack width increases, the critical disorder ratio decreases, as expected.

The peak average stress is used to solve for the critical stress intensity factor as the crack initiates. Therefore, the critical stress intensity factor can be defined as function of the maximum average stress, $\sigma_{\infty,maximum}$, and the crack length, $a$: $K_{IC} = \sigma_{\infty,maximum}\sqrt{\pi a}$ . According to pure elastic arguments, the critical stress intensity factor remains constant as the initial value for $a$ remains constant (Bazant 2004). However, with the contribution of induced-disorder, there is an additional length scale that must be considered. As disorder strength increases, the theory behind notched-specimen fracture mechanics has less of an impact on the crack initiation and propagation behavior. Instead, the crack nucleation and growth behavior is driven by the fluctuating material toughness energy in the phase field model.

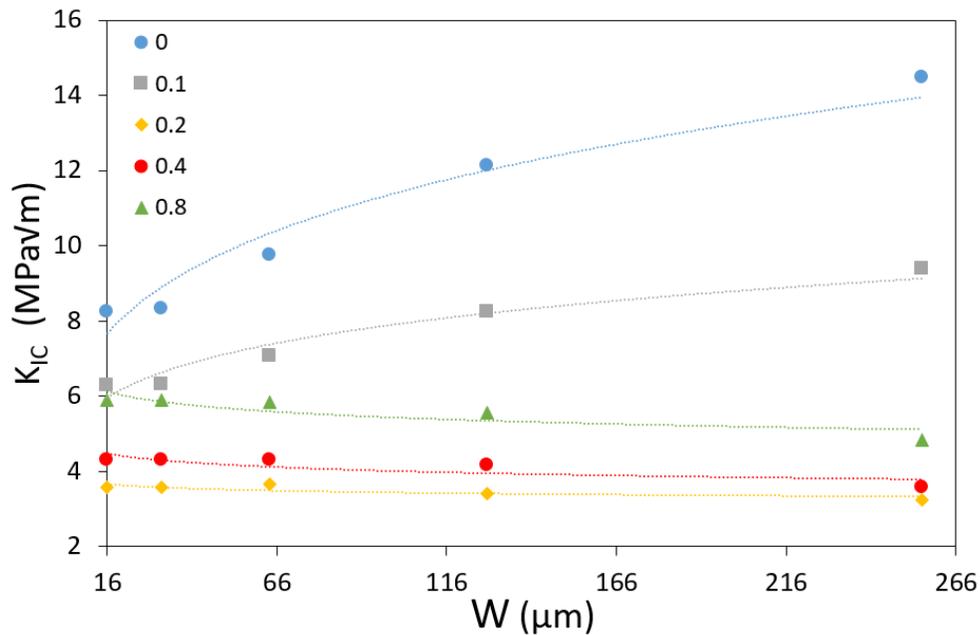

*Figure 10:* **Critical stress intensity as it relates to disorder strength and crack width:** *The critical stress intensity factors*

*at the disorder strengths with respect to notch width, W. Line-guides to the eye are shown for each data set.*

Figure 10 shows the critical stress intensity factor that is the stress intensity factor at fracture as a function of notch width. The behavior of the critical stress intensity factor with the notch width displays a drastically different behavior for small ($R_G < 0.2$) and large ($R_G > 0.2$) relative disorder in the critical strain energy release rate. Namely, for $R_G < 0.2$, $K_{IC}$ displays elastic-like scaling with the notch width. However, as $R_G$ becomes larger than 0.2, the critical stress intensity factor does not scale with the notch width, a signature that disorder-driven crack nucleation has occurred.

Characterizing the effect of increasing the critical strain energy release rate variance across several system sizes is necessary to understanding how fracture behavior transition is impacted by the sample length, $L_y$. One way to indicate this is by the difference in the maximum strain, average strain or other observables at fracture, between the very sharp (W = 16 µm) and very thin notch (W = 0.5 mm) cases. This difference can be viewed as an order parameter for the aforementioned phase transition. As in figure 8 for various notch widths, the average stress curves begin to collapse on each other as the disorder strength increases. Data was collected for three system sizes, ranging from $L_y$ of 0.25 to 1.0 mm. Also, disorder strengths were tested for the range 0.0 to 0.8.

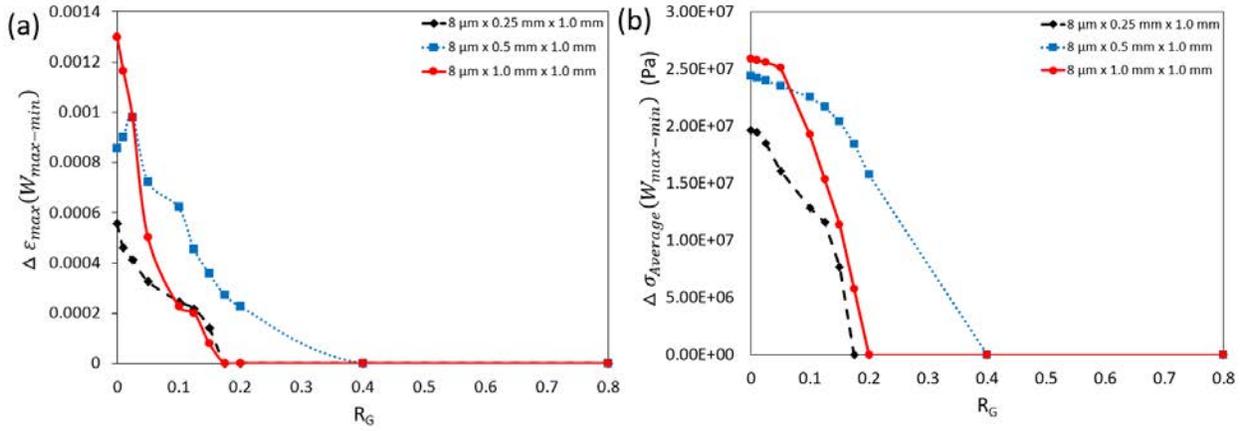

*Figure 11:* **Difference in spatially maximum strain and average stress as a function of disorder strength and maximum and minimum crack width:** *(a) the difference in the maximum strain of the notch widths of 16 µm and 0.5 mm as a function of disorder (b) the difference in the peak average stress of the notch widths of 16 µm and 0.5 mm as a function of disorder.*

As we observed the behavior of the average stress curves (Fig. 8), once the maximum average stress is reached the stress begins to decrease relatively linearly. However, the stress does not obtain a value of zero rather a residual stress value. Figure 11 (a) shows the difference in average strain of the notch width W = 16 µm and 0.5 mm at the location of which the residual stress was present. In panel (b) for both cases of $L_y$ = 0.25 and 1.0 mm, the average stress curve of notch widths 16 µm and 0.5 mm are very similar in behavior and magnitude, collapsing on each other at approximately same disorder strength. The behavior of these curves exhibit a much shaper decline than $L_y$ = 0.5 mm in the maximum difference of average stress for the notch widths. The differences are expected for the fact that disorder is not sample averaged.

For the lower fractal dimension D= 2.85, the critical stress intensity factor values exhibit the same behavioral trend across all disorder strengths. The reason for the absence of an explicit transition is mainly the artificial sharpness of a finitely discretized notch. However, the curves appear to converge as they reach the maximum disorder strength tested.

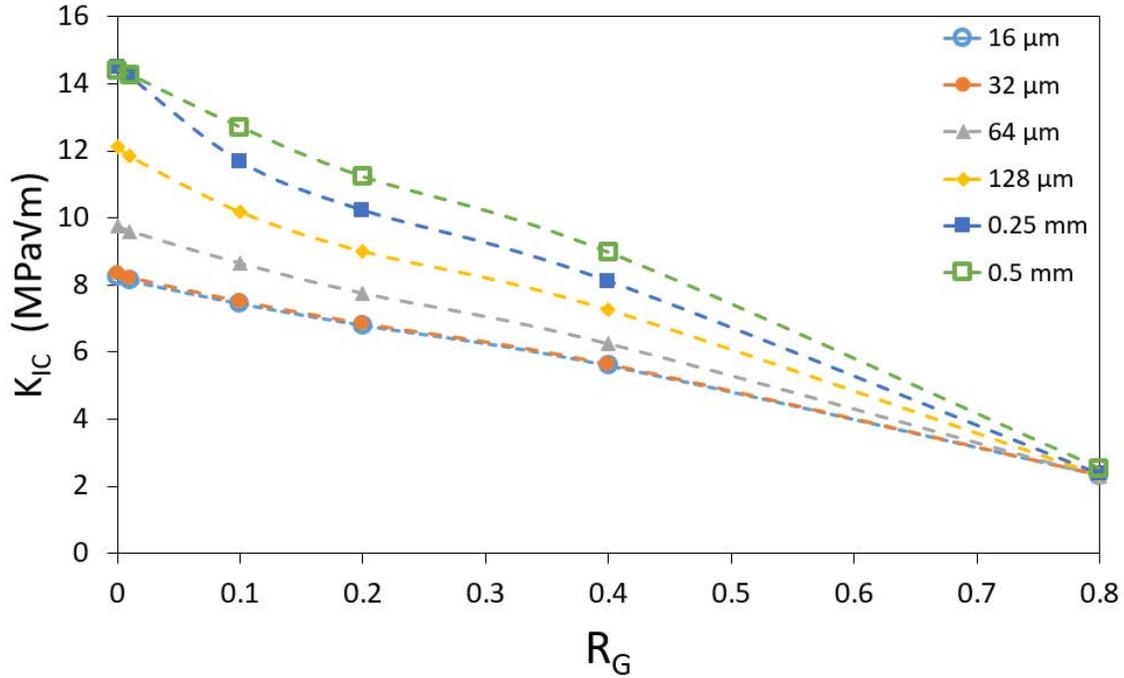

*Figure 12:* **Effect of disorder strength on critical stress intensity factor for fractal dimension D=2.85:** *The critical stress intensity of simulation system size of $L_x$=8 µm, $L_y$=0.25 mm, and $L_z$=1.0 mm where W of 16 µm, 32 µm, 64 µm, 0.125 mm, 0.25 mm, and 0.5 as a function of disorder strength for fractal dimension D = 2.85.*

Figure 12 identifies how the critical stress intensity behaves with respect to the lower fractal dimension D = 2.85. Though the curve appears to converge as the disorder strength is increased, the curves do not cross or overlap. We compare this behavior to the higher fractal dimension D=2.995 where there is an explicit transition at $R_G \sim 0.2$.

The same methodolgy was applied to the critical stress intensity factor of the higher fractal dimension D = 2.995. It can be seen that the transition occurs in the region near a disorder strength of 0.2. The trend is similar to the behavior exhibited in Figure 8.

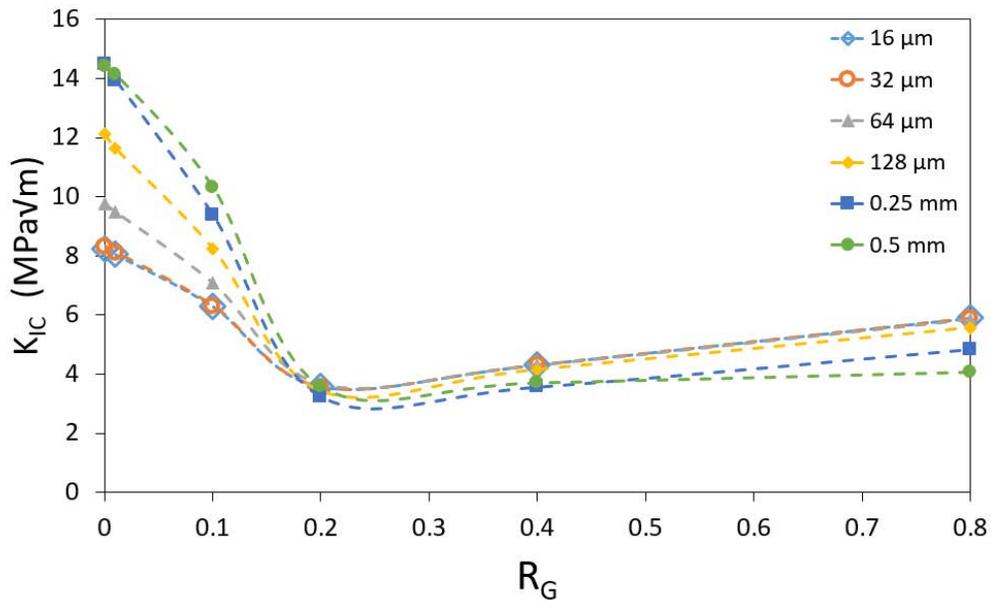

*Figure 13:* **Effect of disorder strength on average stress intensity factor for higher fractal dimension D = 2.995:** *The critical stress intensity of system size of $L_x$=8 µm, $L_y$=0.25 mm, and $L_z$=1.0 mm for W of 16 µm, 32 µm, 64 µm, 0.125 mm, 0.25 mm, and 0.5 mm as a function of disorder strength for the fractal dimension D 2.995.*

In figure 13, the critical stress intensity factor appears to converge at the disorder strength of about 0.2. The critical stress intensity factor after this convergence increases at a slow rate. The radius of curvature at smaller notch widths, like 16 μm and 32 μm, is small enough that the location of the crack initiation will occur almost exclusively at the notch tip in the same way in terms of crack initiation. However, at near zero disorder strength, the notch widths W of 0.25 mm and 0.5 mm exhibit very similar magnitudes of critical stress intensity factor. This eludes to the observation: There is an effective interval where the radius of curvature will have an impact on the critical stress intensity factor at the notch tip i.e. increasing the radius of curvature past this threshold doesn't affect the critical stress intensity factor at zero disorder strength.

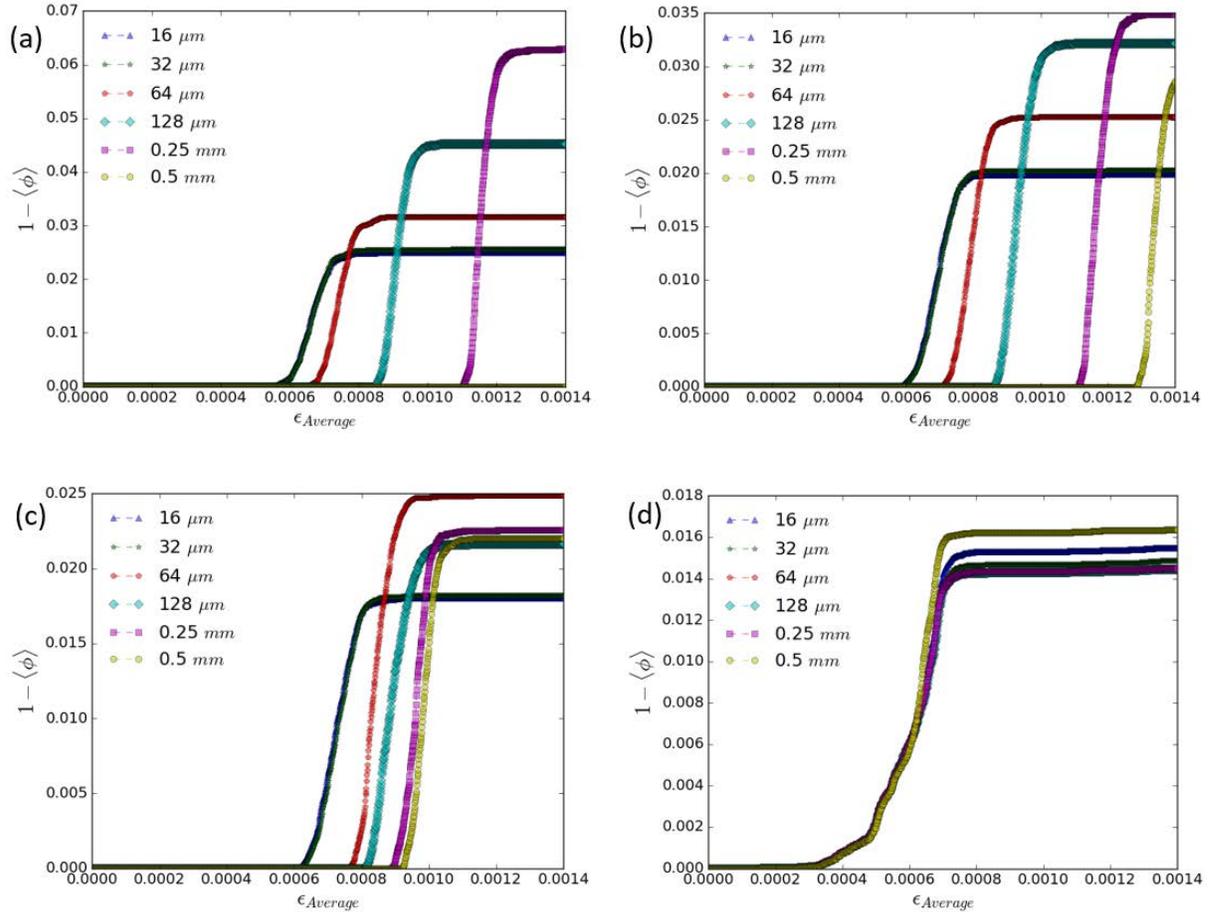

*Figure 14: **Effect of disorder on average damage:** The average damage of the specimens of $L_x$=8 μm, $L_y$=0.5 mm, and $L_z$=1.0 mm where D=3.995 as a function of the average strain with respect to $R_G$ where (1) has a $R_G$ = 0.0, (2) has a $R_G$ = 0.1, (3) has a $R_G$ = 0.2, and (4) has a $R_G$ = 0.4.*

In figure 14, the average damage is plotted as a function of the average strain. However, as damage is a local phenomenon, we track how many elements in the Fourier grid exhibit damage as the displacement-controlled loading case affects the phase field energy. The displacement of the total sample is average and the average strain is what we plot with regards to average damage. The disorder strength increases from 0.0 to 0.4. At lower disorder strengths, the behavior is consistent with the observations made in regards to the average stress where it can be characterized as brittle fracture behavior. Also, at relatively low disorder strengths before the transition, the average damage increases as the notch width increases and the crack initiation occurs at a larger strain. The relationship between maximum achieved average damage and the crack width may be described when considering that the strain value at the crack nucleation for higher crack widths is greater than those in the lower crack width's crack nucleation strain values. For lower disorder strengths, where a single crack initiates at the notch tip, the average damage corresponds proportionally to the crack length as no other damage occurs but at the crack tip as it propagates.

As the disorder strength increases, the behavior of the average damage appears to have converged with respect to the time at which crack nucleation occurs. At $R_G$ = 0.8, the trend seen at the lower disorders experiences a reversal, so as the crack width increases, the maximum average damage achieved decreases. However, typically, there are visible events at higher disorder strengths because of the increased stochastic character of the specimen. With regard to the average damage, there are minor fluctuations in the average damage as it exponentially increases to its final failure point. Furthermore, at the higher disorder strengths, the average damage shows a steady increase as the load case continues even after complete failure. This leads to the assumption that there is some residual stress in the sample that is causing some bonds in the material to break at the point of some of the crack fronts that have not fully propagated through the sample. This transfer in trend only further shows the transition from brittle to quasi-brittle fracture from samples at various disorder strengths.

The rate of change of the average stress curves was analyzed by taking the difference of the increments. The difference was then multiplied by the square of the radius of curvature and then divided by the critical stress intensity factor in order to produce a non-dimensional parameter.

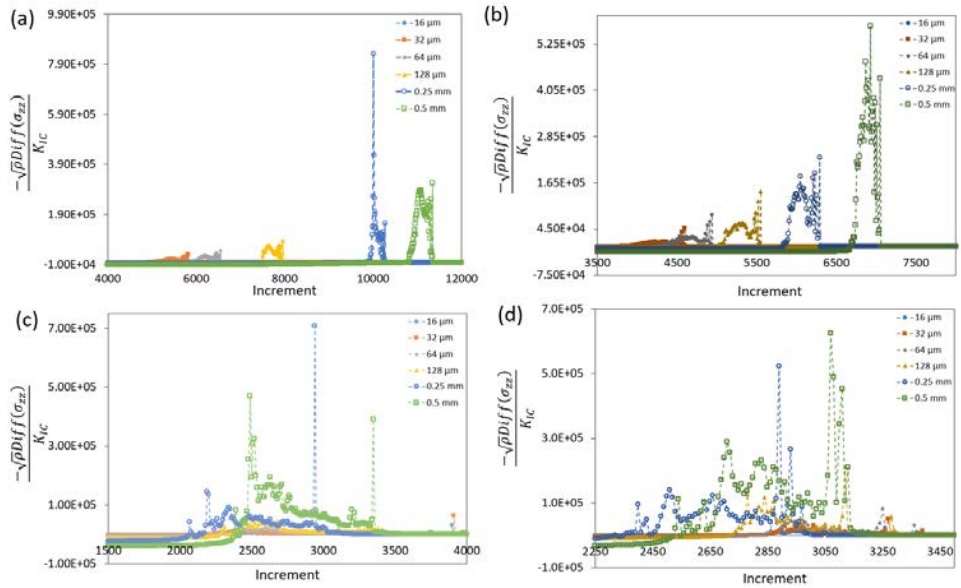

*Figure 15: **Dimensionless, instantaneous rate of change of the average stress:** The dimensionless term for the instantaneous rate of change of average stress of system sizes $L_x$=8 μm, $L_y$=0.25 mm, and $L_z$=1.0 mm where D=2.995 as a function of the incremental step with respect to $R_G$ where (a) has a $R_G$ = 0.0, (b) has a $R_G$ = 0.1, (c) has a $R_G$ = 0.2, and (d) has a $R_G$ = 0.4.*

Figure 15 shows the plots of the non-dimensional parameter for the rate of change of the average stress versus incremental step. Since the simulations are displacement controlled, the behavior of the plots remains unaltered when considering an increment of the average strain across the sample. Fundamentally, the derivative of the average stress plots would show a constant value, initially, as the constant slope of elastic stress remains consistent with the increasing load case. However, after complete failure, the rate of change in the average stress exhibited across all simulations is large. The sharp increase in the rate of change of average stress occurs at about a magnitude of $10^2$ kPa before the parameter becomes dimensionless. This observation indicates that the crack has fully propagated through the sample and the average stress has dropped to values at which it is considered the residual stress. As the disorder strength increases, the event of complete failure across all crack widths begins earlier in the load case. In panel (a), the rate of change in average stress appears to exhibit the same general trend as the average stress curve. As the disorder strength is increased, the change in the average stress begins to exhibit the same stochastic behavior of the respective disorder strengths.

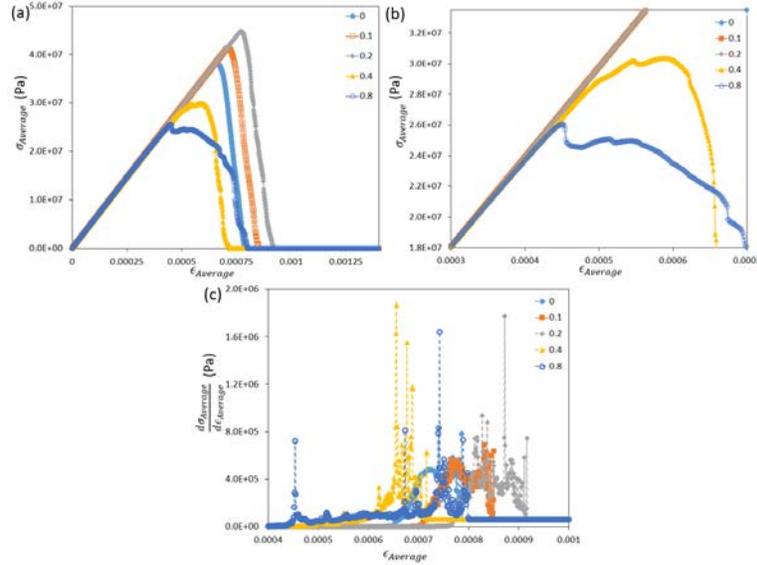

*Figure 16: **Quantification of the fluctuations at crack nucleation and growth of average stress:** Simulations of sample size $L_x$=8 μm, $L_y$=0.5 mm, and $L_z$=1.0 mm where D=2.995 and W = 64 μm as a function of the average strain where (a) is the average stress vs. average strain, (b) is the zoomed-in view of the fluctuations for disorder strengths 0.4 and 0.8, and (c) is the difference in the rate of change of the average stress and the initial rate of change of linear increase in average stress vs. average strain.*

These abrupt crack-initiation events correspond to avalanche precursors to crack initiation in the model we investigated. The precursors become larger and more intricate as $R_G$ increases, and may amount to 1-5MPa for aluminum (Fig.16), a clearly observable range in typical experimental set-ups. However, in the dimensionless parameters, these avalanche precursors' magnitude appear to remain invariant with $R_G$, suggesting that they may become useful for the prediction of generic crack initiation in intermetallics and other alloys (Ritchie et al. 2001).

## 4    Conclusions

A simulation-based model of aluminum is developed to obtain a better understanding of the realistic response that a specimen would undergo in mode I fracture failure. Analytically, most materials are assumed to be homogenous with constant material properties throughout the sample. However, realistically, through manufacturing and processing, any sample of a material has fluctuations in material properties at the microscale. We considered this assertion valid because if we engineer failure in several alloys under the same loading conditions, then the fracture characteristics will be slightly different from sample to sample. Also, conducting multiple trials in an experiment implies that the behavior of geometrically identical samples will fracture differently.

The Weierstrass- Mandelbrot function was implemented in a material simulation software, DAMASK (Roters et al. 2012), towards the investigation of crack initiation in disordered microstructures that may deform elastically and plastically. A phase field model was implemented to control the critical strain energy release rate of each specimen. Our modeling methods are powerful in that they can carefully describe micromechanical material properties, in a constitutive manner, such as viscoplasticity, hardening, twinning, elastic anisotropy, and they may be used to simulate the fracture behavior. The fractal dimension of the stochastic, microstructural properties, D, is increased from 2.85 to 2.995 to increase the magnitude of the critical strain energy release rate fluctuations, in comparison to the artificial magnification of stress fluctuations induced by the modeling resolution of the material geometry. Physically, D is identified as characterizing the fractal surface geometry of the material property of the critical strain energy release rate. The initial crack width was varied from 16 μm to 0.5 mm by powers of 2. For brittle fracture, the theory of fracture mechanics expects that crack initiation would occur at the notch tip. But, this behavior is, also, exhibited at lower disorder strengths where the critical strain energy release rate variance is not large enough to have an effect on the initial crack nucleation location. At higher disorder strengths, the critical strain energy release rate variance is large enough to induce crack nucleation on the bulk sample where the stresses are higher than those present at the notch tip.

The purpose of increasing the disorder strength in the material simulations is to identify the transition of brittle to quasi-brittle fracture behavior. When examining the average stress data, the lower disorder strength and small notch width simulations exhibit behavior more closely characterized as brittle fracture where there were no large or abrupt changes in the average stress curves. At lower disorder strength, but larger initial crack widths, the rate of change in the average stress begins to increase on the scale of 100 kPa. At higher disorder strengths, all initial crack widths specimens break at relatively the same point but the curves collapse on each other. At lower disorder strengths, the peak average damage of the specimen increases as notch width increases, while at higher disorder strengths, the curves across all notch widths begin to collapse on top of each other, exhibiting the same general behavior as the average and maximum stress curves. The maximum value of the maximum stress plots across all of the simulations is approximately 200 MPa. All of the specimens exhibit similar maximum stress values because the yield stress and average critical strain energy release rate in the simulations remains constant throughout the simulations. This relation between notch-driven and disorder-driven fracture is identified through the behavior during crack initiation and quantified from parameters like average stress, maximum stress, and average damage.

In the future, the aim is to test several samples of different manufacturing and processing background and characterize the degree of disorder based on the material and fracture behavior. This paper shows that there is a relationship of the fracture behavior between the notch-driven (geometry-based) and disorder-driven (material-based) fracture where it transitions from brittle to quasi-brittle fracture behaviors for crack nucleation and propagation. These simulations demonstrate that there are observable avalanche precursors that could be tracked in experimental efforts through well controlled experimental setups by just regulating the geometry of the notch on 2D samples. It is plausible that the magnitude of these avalanche precursors extend into a non-trivial scaling function that can be used throughout the critical regime; if that's the case, as this work suggests, then it should become possible to utilize these events towards safe and non-invasive prediction of crack initiation in intermetallics and other disordered brittle materials. Finally, it is important to extend this study into three dimensions and also to ductile fracture, which are straightforward steps, given our current modeling flexibilities.

## 5    Acknowledgements

We would like to thank M. Alava, C. Reichhardt, H. Song, S. Zapperi and E. Van der Giessen for helpful and inspiring comments on this manuscript. We would also like to thank J. El Awady and K. Hemker for inspiring discussions in the beginning of this project. We would like to acknowledge the use of the WVU High Performance Computing facilities for this project. Furthermore, we would like to acknowledge support from UES under subcontract S992009001 from AFOSR (SP) in a part of this project.